\documentclass{elsart}
\usepackage{graphicx}% Include figure files
\usepackage{natbib}
\usepackage{amssymb}

\begin{document} 
\begin{frontmatter}
\title{Log-Poisson Statistics and Extended Self-Similarity in Driven Dissipative Systems}
\author[Singapore]{Kan Chen\corauthref{cor1}} 
\ead{kan\_chen@nus.edu.sg}
\ead[url]{http://www.cz3.nus.edu.sg/$\sim$chenk}
and
\author[OSU]{C. Jayaprakash}

\address[Singapore]{
Department  of  Computational Science, Faculty of Science
National University  of Singapore, Singapore 117543
}
\corauth[cor1]{Corresponding author.}

\address[OSU]{Department  of  Physics, The Ohio State University,
Columbus, OH 43210, U.S.A.
}

\begin{abstract}

  The Bak-Chen-Tang forest fire model \cite{BCT} was proposed as a toy model of turbulent systems,
  where energy (in the form of trees) is injected uniformly and globally, but is dissipated (burns)
  locally.  We review our previous results on the model \cite{CB,BC} and present our new results
  on the statistics of  the higher-order moments for the spatial distribution of fires.  We show
  numerically that the spatial distribution of dissipation can be described by  
Log-Poisson statistics  which leads to extended self-similarity (ESS) \cite{Benzi93,Dubrulle}. 
Similar behavior  is also found in models based on directed percolation; this suggests 
that the concept of Log-Poisson statistics of (appropriately normalized) variables can be 
used to  describe scaling not only in  turbulence but also in a wide range of driven dissipative systems.

\end{abstract}


\end{frontmatter}

\section{Introduction}

Understanding the spatial distribution of energy dissipation is
crucial in the study of  homogenous, isotropic, fluid turbulence.
Kolmogorov, in his celebrated 1941 theory (K41)~\cite{K41}, first
conjectured that, in the inertial range, the velocity structure
functions scale as a power law in the separation $l$. The theory
implicitly assumes uniform energy dissipation characterized by a
mean dissipation rate. Experiments from the recent decades showed
clear deviations from the K41 scaling, which are believed to be
related to the highly intermittent and non-uniform spatial
distribution of dissipation.~\cite{frisch,sreeni} Many
phenomenological models of energy cascade and intermittency have
been proposed to understand the deviations from the K41 scaling.
The most famous are the Lognormal model~\cite{K62}, the
Multifractal model~\cite{Parisi}, and the Log-Poisson model by She
and L\'{e}v\^{e}que (SL)~\cite{She}. The SL model, based on a hypothesis about the behavior of the
moments of the energy dissipation, leads to a prediction of the
exponents for the velocity structure functions that appears to be
in excellent agreement with experimental results.  Benzi et al.~\cite{Benzi93}
  discovered an important property they termed
Extended Self-Similarity (ESS): the $p^{th}$ order structure
function has a power-law dependence on the third-order structure
function. They found that the validity of  ESS extended almost
down to the dissipative range, roughly five times the Kolmogorov
scale. The connection between the dissipation distribution and the
scaling in the inertial range is further supported by these
investigations.

It remains  a great challenge to understand the dynamical
mechanism for the inertial range scaling and
 ESS in turbulence. So far it has not been feasible to attack the problem directly from
the underlying equations for turbulence. Instead we choose to
study a few simple models of ``turbulent systems", such as the BCT
forest fire model. We hope that, by illustrating the emergence of
complex spatial distribution of dissipation and ESS in such simple
dynamical models, we can gain some intuition about the dynamical
mechanism for energy cascade and scaling in fully developed
turbulence.

\section{Scale-Dependent Dimension in the BCT Forest Fire Model}

The BCT forest fire model is defined as follows. On a
$d$-dimensional lattice of size $L^d$, trees, representing energy,
are grown randomly at empty sites at a rate $p$. At each time
step, trees that are on fire are burnt down (the site becomes
empty at the next time step) and neighboring trees are ignited.
The fires die out when there are no more trees in their
neighborhood. These processes represent the dynamics of energy
input and dissipation. Numerical studies indicate that there is a
correlation length $\xi$ ($\xi \sim p^{-2/3}$ in 3D) in the system
\cite{CB}. If $L\leq \xi$ the fires cannot be sustained, so we
consider only the case $L > \xi$. Larger and larger systems can be
studied in the limit $p \rightarrow 0$. Despite the simplicity of
the model, many fascinating scaling behaviors emerge. For example,
Br\"{o}ker and Grassberger~\cite{grass} found anomalous scaling in
three and four dimensions; in particular, they found periodic
oscillations in autocorrelation functions with a period $T$ which
diverges as $1/p^\beta$ with $\beta\,\approx\,0.77$ in three
dimensions.

\begin{figure}
\includegraphics*[width=12cm]{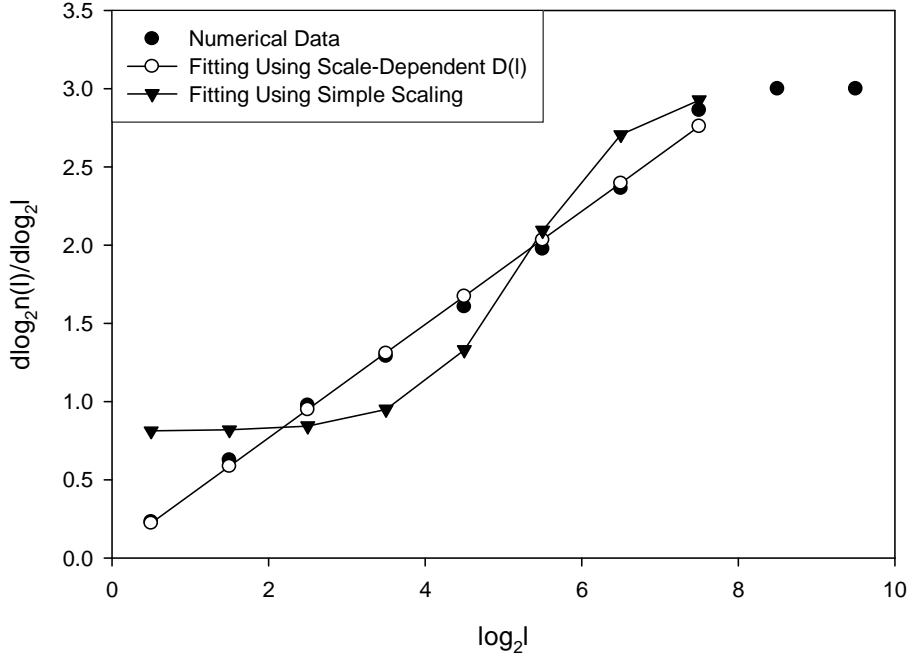}
\caption{ $D(l)$ vs $\log_2 l$ for the BCT forest fire model with
$L=1024$ and $p=0.0003$. Averages were obtained using $10^6$ time
steps (after skipping $200,000$ time steps).  The fits using the
expression of scale-dependent dimension and the simple ``fractal"
scaling with crossover to 3D are also shown.}
\end{figure}

One of the interesting features of the model is the unconventional
scaling  that can be interpreted geometrically as a
scale-dependent fractal dimension: the apparent dimension $D(l)$
characterizing the distribution of fires at the length scale $l$
in three spatial dimensions increases linearly in 
$\log(l)$ \cite{CB,BC}. The dimension increases gradually as one
steps further and further backwards, and views the system at
larger and larger length scales. At some point, the correlation
length is reached; the distribution of fires remains uniform
beyond that length. This is the main geometrical picture for the
spatial distribution of the fires. We have repeated the simulation
for larger lattices (up to $2048^3$) and smaller $p$ (as small as
$0.00015$); so far no deviation from the scaling described above
is found. As was done in the previous simulation \cite{CB}, we
determine the mean number of fires $n(l)$ within boxes of size $l
< \xi$ that contain fires. The linear dependence of $D(l)$ on
$\log(l)$ corresponds to
\begin{equation}
\log(n(l)) \sim \alpha \frac{\log(l/l_0)}{\log(\xi/l_0)} \log(l/l_0),
\end{equation}
which gives rise to
\begin{equation}
D(l) \equiv \frac{d\log(n(l))}{d\log(l))} = 2\alpha
\frac{\log(l/l_0)}{\log(\xi/l_0)} \label{dl}\,.
\end{equation}

Fig.~1 shows the numerical values of $D(l)$  vs $\log_2 l$. Here
$D(l)$ is calculated using the numerical derivative: $D(2^{1/2} l)
=  \log_2(n(2l)) - \log_2(n(l))$. In the figure we also show the
two-parameter fit using the above equation. There is excellent
agreement between the data and the our proposed expression of
$D(l)$ (even a one-parameter fitting with $l_0$ fixed to $l_0 =1$
is very good). For comparison we also show the fitting using a
simple ``fractal'' scaling: $n(l) = a l^b + c l^3$ which gives
$D(l) = (abl^b + 3cl^3)/(al^b + cl^3)$. This three-parameter
fitting is very bad. It is interesting to note that the
distribution of luminous matter in the universe might also follow
a  scaling form similar to the one given in Eq.~\ref{dl}
\cite{BC}.

\section{Extended Self-Similarity in the Forest Fire Model}

The scale dependent dimension shown in the previous section only concerns the first moment of the fire
distribution. We will show that the higher-order moments also exhibit interesting scaling behaviors.
Preliminary study in Ref.~\cite{CB1} indicated that the higher-order moments can be described by extended
self-similarity  used to describe the moments of energy dissipation in turbulence. We present here a
detailed study of the higher-order moments.

\begin{figure}
\includegraphics*[width=12cm]{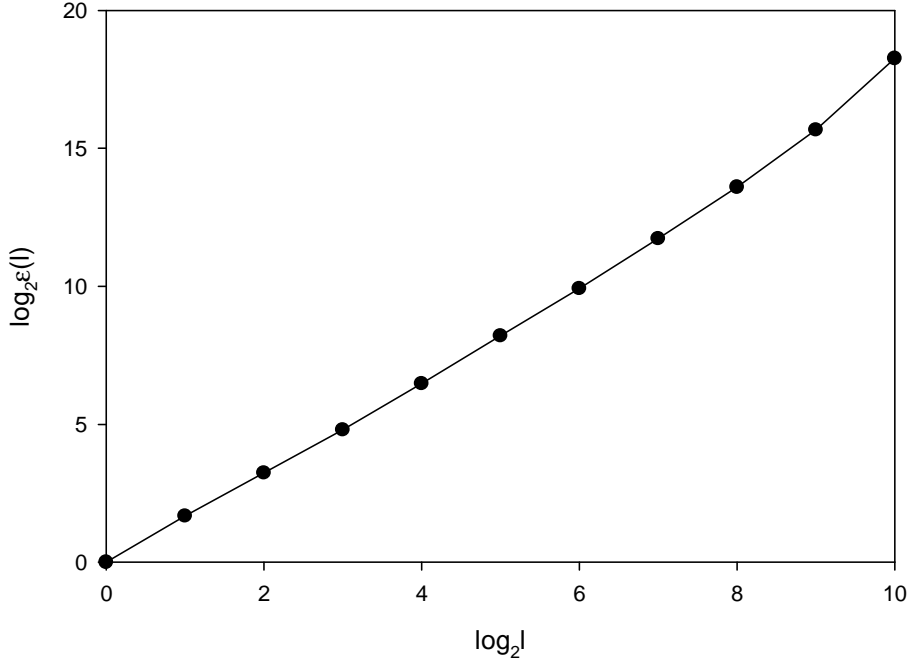}
\caption{ $\log_2\epsilon_{\infty}(l)$ vs $\log_2 l$ for the BCT forest fire model with $L=1024$ and
$p=0.0003$}
\end{figure}

We start with the higher-order moments given by
\begin{equation}
M_q(l) = <m(l)^q>,
\end{equation}
where $m(l)$ is the number of fires contained in a box of linear
size $l$; we emphasize that the average is again over all the
boxes that contain fires. Following  Dubrulle's analysis of
generalized scale covariance and ESS in
turbulence~\cite{Dubrulle}, we consider the normalized variables, defined by $\pi(l)=m(l)/\epsilon_{\infty} (l)$, where 
$\epsilon_{\infty}(l)$ given by
\begin{equation}
\epsilon_{\infty}(l) = \lim_{q\rightarrow \infty} \frac{M_{q+1}(l)}{M_q(l)}.
\end{equation}
We focus on these normalized variables and calculate their moments, defined as
\begin{equation}
\Pi_q(l)=<\pi(l)^q>.
\end{equation}
We will show numerically that the scaling behavior of $\{\Pi_q\}$
is consistent with the Log-Poisson statistics. Let $\pi(l) =
\beta^j$. If $j$ is a random variable described by a Poisson
distribution $P_l(j)=\frac{\lambda_l^j e^{-\lambda_l}}{j!}$, then
$\pi(l)$ is described by a Log-Poisson distribution. The moments
of $\pi(l)$  can then be written as
\begin{equation}
\Pi_q(l) = \sum_{j=0} \beta^{qj} \frac{\lambda_l^j e^{-\lambda_l}}{j!}
    = \exp[\lambda_l(\beta^q-1)].
\end{equation}
It is easy to see that the moments from this Log-Poisson
distribution are related by
\begin{equation}
\Pi_q(l) = \Pi_p(l)^{\zeta(q)/\zeta(p)}=\Pi_1(l)^{\zeta(q)},
\label{piq}
\end{equation}
where 
\begin{equation}
\zeta(q) = (1-\beta^q)/(1-\beta).
\label{zeta}
\end{equation}
Note that $\zeta(q)$ is not linear in $q$ as would be the case for simple scaling.
 The moments of the original variables $m(l)$ are given by
\begin{equation}
M_q(l)=\epsilon_{\infty}^{q}(l)\Pi_1(l)^{\zeta(q)}.
\end{equation}

To test whether the moments can be described by the Log-Poisson distribution numerically, we first fit
the data $M_q(l)$ to the above expression. The actual fitting is done using the following expression:
\begin{equation}
\log_2(M_{q+1}(l)/M_q(l))=a(l) + b(l) \beta^q,
\end{equation}
where $a(l)=\log_2\epsilon_{\infty}(l)$ and $b(l) =
\log_2(\Pi_1(l))$. We first determine $\beta$ as follows: For each
value of $l$ we do a linear fitting of $\log_2(M_{q+1}(l)/M_q(l))$
vs $\beta^q$, and calculate the overall error, which depends on
$\beta$. The best value of $\beta$, corresponding to the minimum
overall error, is found to be $\beta \approx 0.75$. With $\beta$
fixed to this value, we find $a(l)$ or $\epsilon_{\infty}(l)$ from
the fitting; this is plotted in Fig.~2.

\begin{figure}
\includegraphics*[width=12cm]{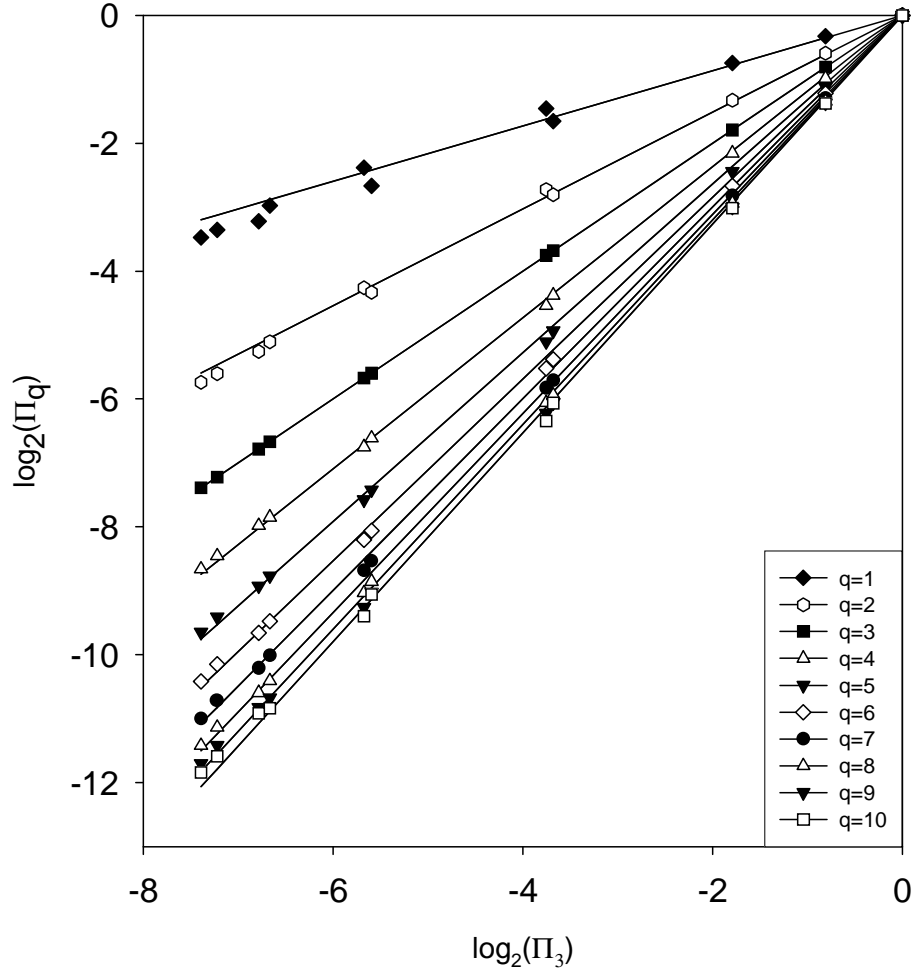}
\caption{ $\Pi_q$ vs $\Pi_3$. The lines are drawn with slopes $\zeta_q/\zeta_3$, where $\zeta_q$ are
defined in Eq.~\ref{zeta}. The value of $\beta=0.75$ is chosen as explained in the text.}
\end{figure}

For $l<\xi$ ($\xi$ is about $2^8$ for the value of the parameter
$p=0.0003$ used in the simulation), $\epsilon_{\infty}(l) \sim
l^{1.8}$. Note that if the original variable $m(l)$ is defined as the density of fires (
instead of the number of fires) in a box of linear size $l$, $\epsilon_{\infty}(l)$ will be approximately
given by $l^{1.8}l^{-3} =l^{-1.2}$ --- but the normalized variables remain the same. 
Given $\epsilon_{\infty}(l)$ we can evaluate the
normalized moments $\Pi_q(l)$. If the distribution is Log-Poisson,
then $\Pi_q$ vs $\Pi_p$ in the log-log plot should fall on a
straight line with a slope given by $\zeta(q)/\zeta(p)$. Fig.~3
shows $\Pi_q$ vs $\Pi_3$ for $q=1,2,\cdots,10$. 

Indeed, we find
straight lines in the log-log plots with slopes matching
$\zeta(q)/\zeta(3)$. This is extended self-similarity (in the
normalized variables) for the forest fire model. This is
particularly interesting, given the fact that there is actually no
scaling regime in the plot of $M_q(l)$ or $\Pi_q(l)$ vs $l$. The
Log-Poisson distribution provides an excellent description of the
normalized moments.

\section{Extended Self-Similarity in Models of Directed Percolation}

\begin{figure}
\includegraphics*[width=12cm]{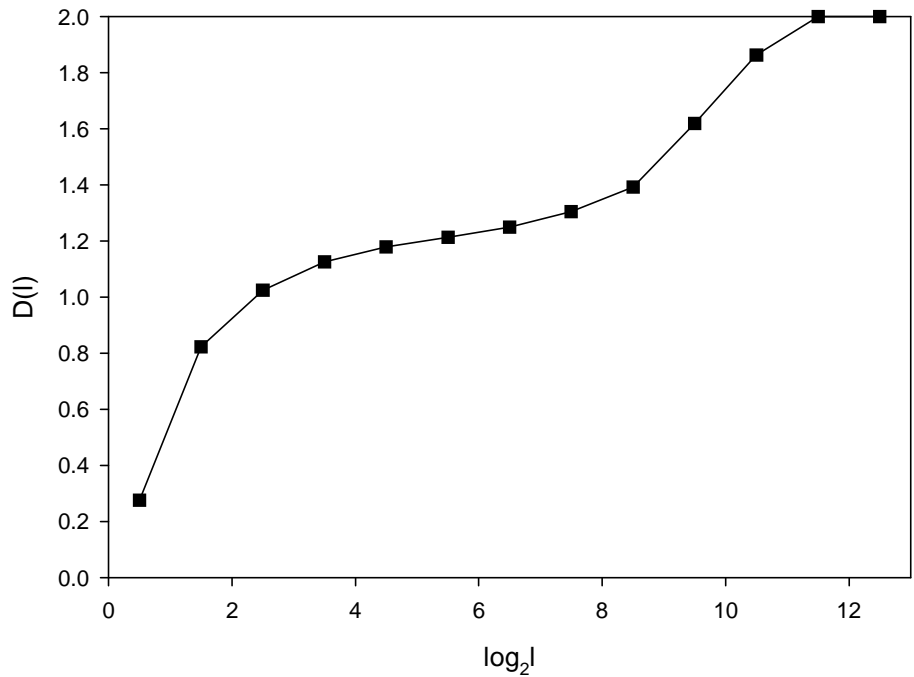}
\caption{ $D(l)$ vs $\log_2 l$ for $2+1$ DP model with $L=8192$ and $p=0.28734$. $5\times10^5$ time
steps (after skipping $100,000$ time steps) were used to obtain the averages.}
\end{figure}

It is of interest to determine whether the two properties of the
forest fire model described in the previous section are displayed
by other models.  The first, the special form of the
scale-dependent dimension  appears to be unique to the forest fire
model; so far we have not found other types of models which
exhibit similar scaling in the first moment. However, the second
property of extended self-similarity in normalized variables as
described in the last section appears to be much more general.
Since the fire propagates and spreads in the background of trees
which has presumably a hierarchical and certainly a dynamically
changing topology it is interesting to consider simpler models
which describe a spreading process.  The most obvious one is the
well-studied model of directed percolation (DP).~\cite{dp} We
found to our surprise that quite a few models of directed
percolation (DP) and their variants exhibit a similar form of ESS
based on log-Poisson statistics.  As an example we consider $2+1$
DP in the supercritical phase $p>p_c$, which is very different
from the 3d forest fire model. Fig.~4 shows $D(l)$ vs $l$ for this
model. Here we use a lattice of size $8192 \times 8192$ and
$p=0.28734$. Over half millions time steps were used to obtain the
averages.

\begin{figure}
\includegraphics*[width=12cm]{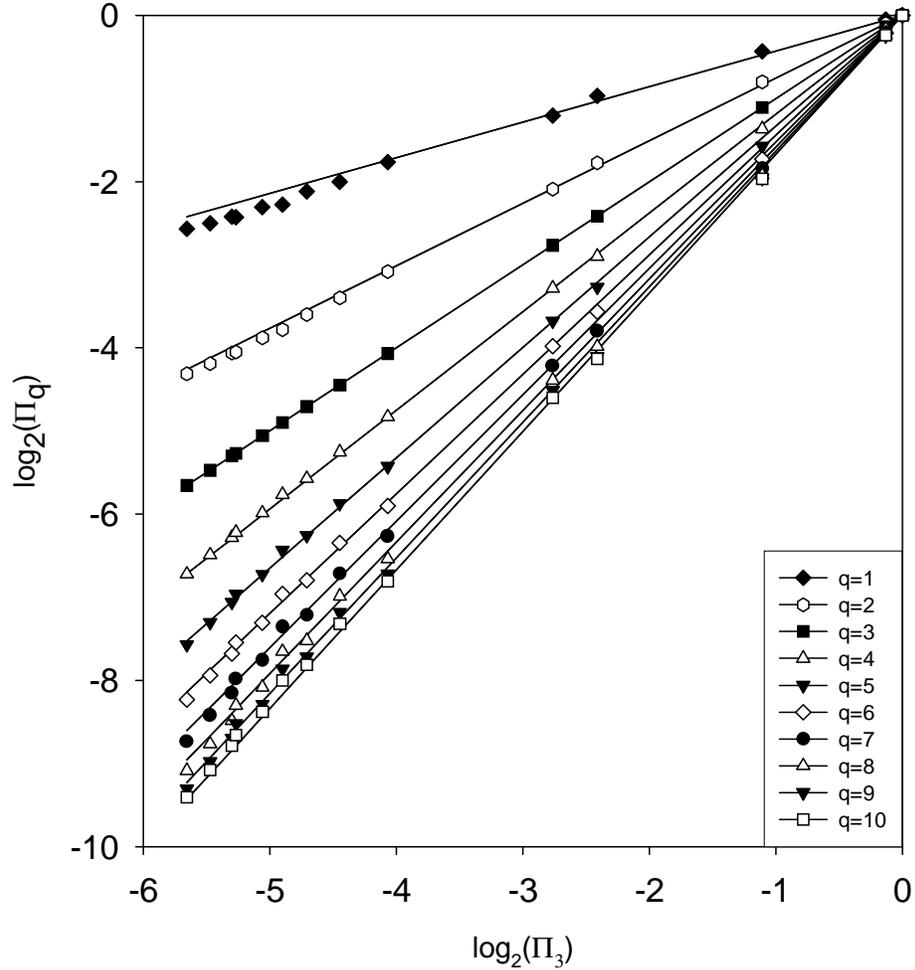}
\caption{ $\Pi_q$ vs $\Pi_3$ for $2+1$ DP model. The lines are drawn with slopes $\zeta_q/\zeta_3$, 
where $\zeta_q$ are defined in Eq.~\ref{zeta}. The value of $\beta=0.76$ is used.}
\end{figure}

There is an approximately scale-invariant regime corresponding to $D(l)$
being constant, in contrast to the 3d forest fire model. However,
the normalized higher order moments can still be described the
Log-Poisson statistics (with $\zeta_q$ defined in Eq.~\ref{zeta}) 
and the associated ESS, as can be seen from
Fig.~5. This data provides evidence that either the model studied
or a model in its vicinity in parameter space exhibits the
behavior displayed asymptotically; the proximity to the
appropriate fixed point controls this behavior and elucidating its
features would be of interest.

\section{Summary}

We presented extensive numerical evidence to show that the BCT
forest fire model exhibits two interesting properties: a
logarithmic scale-dependent dimension and  Log-Poisson
distribution of normalized variables (with $\zeta_q$ not linear in
$q$) and associated extended self-similarity. The former is likely
a unique feature of the forest fire and closely related models;
the latter, however, appears to be more general. We have shown
that models based on directed percolation also exhibit Log-Poisson
statistics (for appropriately normalized variables) similar to
that in the forest fire model. This suggests that the Log-Poisson
distribution and the associated ESS can be used to describe
scaling in a wide range of driven dissipative systems. We hope
that studying simple models such as the forest fire model will
provide us with some intuition and physical pictures that might
eventually be useful for understanding the statistics of energy
dissipation and scaling in fully developed turbulence.

This work was supported by the National University of Singapore
research grant R-151-000-028-112. KC thanks Maya Paczuski, Peter
Grassberger, and Itamar Procaccia for helpful discussions. This
work was originally initiated by Per Bak, who was very interested
in scaling and ESS in turbulence. He had hoped that a physical
picture for ESS could emerge in a simple dynamical model such as
the forest fire model. It is sad that he is not around to offer
his unique insight on these topics. For many, particularly those
who worked closely with him, his premature death has left a big
void.

\end{document}